\documentclass{ws-procs9x6}

\begin{document}

\title{Timelike Form Factors}

\author{Kamal K. Seth}

\address{Northwestern University,\\
Evanston, IL 60208, USA\\
$^*$E-mail: kseth@northwestern.edu}

\begin{abstract}
Form factors of nucleons and mesons with timelike momentum transfers are discussed.  New experimental results for protons, pions, and kaons at large momentum transfers are presented, and the inadequacy of existing theoretical ideas about these is pointed out.
\end{abstract}


\bodymatter

\section{Introduction}

It is exactly 100 years that the proton (more correctly, the positive nucleus) was discovered, and 60 years since the pion and kaon were identified.  Since 1964, when the quark model was first proposed, we have known that baryons and mesons are all made up of quarks and antiquarks.  One would therefore expect that by now we could have figured out exactly how quarks fit into the hadrons and lead to their observable properties, mass, size, spin, charges and curents.  But Nature is much more devious, and does not allow easy insight into its workings.  One of the tools that has been successfully used to gain insight into the structure of hadrons is the measurement of electromagnetic form factors as a function of momentum transfer.

Electromagnetic form factors of a hadron are the most direct link to the structure of the hadron in terms of its constituents.  They describe the coupling of a photon with a certain four--momentum to the distribution of charges and currents in the hadron.

The four--momentum transfer $Q^2$ in the collision of two particles with four-momenta $p_1$ and $p_2$ can be positive or space-like (in scattering) or negative or time-like (in annihilation/production).

No understanding of form factors can be considered complete unless it includes an explanation of form factors for both spacelike and timelike momentum transfers, which are just two sides of the same coin.  

\begin{figure}[!tb]
\begin{center}
\hspace*{-0.8cm}
\begin{tabular}{cc}
\includegraphics[width=1.8in]{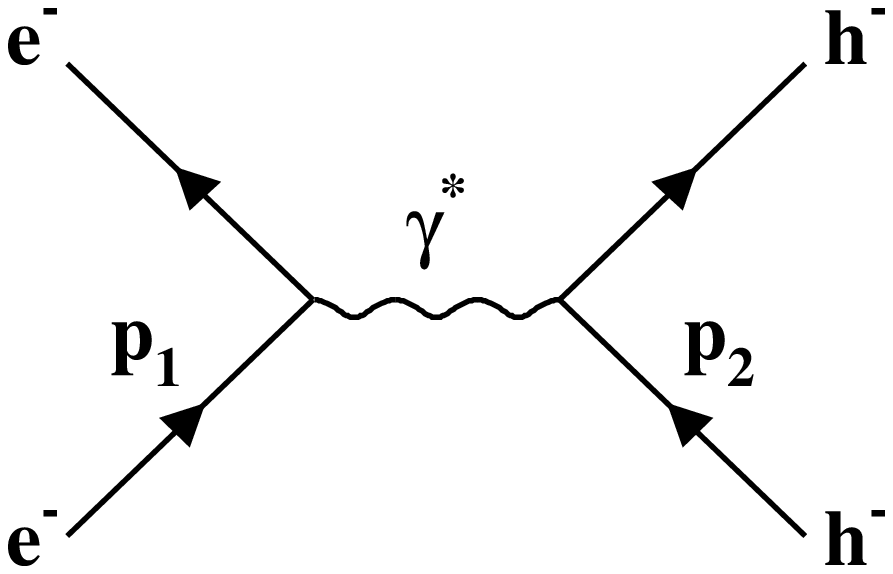}
\vspace*{-18pt}
&
\includegraphics[width=1.8in]{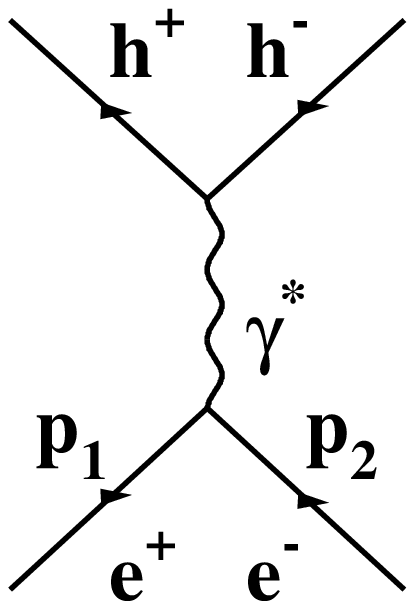}
\\
Scattering, Spacelike
&
Annihilation, Production
\\
positive $Q^2=t$
&
negative $Q^2=s$
\end{tabular}
\end{center}
\end{figure}

\subsection{Form Factors for Space-like Momentum Transfers}

The extensive nucleon form factor measurements done at SLAC, JLab, and other electron accelerators are made with electron beams elastically scattered from targets of $p$, $d$, etc., and are exclusively for spacelike momentum transfers.  Measurements of spacelike form factors of mesons at large $Q^2$ are extremely difficult, if not impossible to make because meson targets do not exist!  Measurements by means of either scattering of pion and kaon beams with atomic electrons, or by electroproduction of pions are largely confined to small momentum transfers

\subsection{Form Factors for Time-like Momentum Transfers}

Measurements of form factors for time-like momentum transfers are done at $e^+e^-$ colliders, and they can, in principle, be used to measure form factors of any mesons or baryons.  The $p\bar{p}$ annihilations have so far been only done with $\bar{p}$ beams incident on fixed proton targets.  These, of course, only lead to proton form factors.

It is important to note that form factors are analytic functions of $Q^2$.  Therefore, the Cauchy theorem alone guarantees that $F(Q^2,\textrm{timelike}) \stackrel{\atop Q^2\to\infty}{\longrightarrow} F(Q^2,\textrm{spacelike})$.

\subsection{Cross Sections for Time-like Momentum Transfers}

For protons, there are two form factors, Pauli and Dirac Form Factors, or more familiarly, the magnetic $G_M(s)$ and the electric $G_E(s)$ form factors, and the cross section $e^+e^-\to p\bar{p}$ is
$$\sigma_0(s) = \frac{4\pi\alpha^2}{3s}\beta_p \left[ |G_M^p(s)|^2 + \frac{\tau}{2}|G^p_E(s)|^2\right],~~\tau\equiv4m_p^2/s$$

At large momentum transfers separation between $G_M(s)$ and $G_E(s)$ becomes difficult, and the results which are generally reported assume $G_E(s)=0$, or $G_E(s)=G_M(s)$.

For pions and kaons, both of which have spin 0, there is no magnetic contribution, and only the electric form factor $F(s)$ exists.  In this case the cross section for $e^+e^-\to m^+m^-$ is
$$\sigma_0(s) = \frac{\pi\alpha^2}{3s}\beta_m^3|F_m(s)|^2$$

Jumping the gun a little, let me point out that pQCD counting rules predict that the baryon form factors are proportional to $Q^{-4}$ (or $s^{-2}$) and the meson form factors are proportional to $Q^{-2}$ (or $s^{-1}$), so that
$$ \sigma_0(s)_{\mathrm{proton}} \propto s^{-5}, \qquad \sigma_0(s)_{\mathrm{meson}} \propto s^{-3}$$
This tells you how rapidly the cross sections fall, and how difficult it becomes to measure any form factors at large momentum transfers.

For example, $\sigma(e^+e^-\to p\bar{p})\approx1~\mathrm{pb}$ at $s=Q^2=13.5~\mathrm{GeV}^2$.  At $s=25~\mathrm{GeV}^2$ one expects to drop down by a factor $\sim20$, to $\sim50~\mathrm{fb}$!

\section{Baryon Form Factors}

The only baryon form factors that have ever been measured are for nucleons, mainly for the proton.  Timelike form factors can also be measured for other baryons, $\Lambda$, $\Sigma$, etc., but so far no such measurements exist.

\subsection{Spacelike Form Factors of the Proton}

The spacelike magnetic form factors $G_M(Q^2)$ of the proton were measured with precision in the ep scattering experiments at SLAC, all the way up to $Q^2=31~\mathrm{GeV}^2$ \cite{slac1}.  For $Q^2\ge15~\mathrm{GeV}^2$, their variation follows the pQCD counting rule prediction that $Q^4G_M(|Q^2|)/\mu_p$ is essentially constant and varies only as $\alpha^2(\mathrm{strong})$.

In the pQCD factorization formalism of Brodsky and Lepage \cite{brodsky}, the form factor can be factorized into the hard scattering amplitude, which can be calculated perturbatively, and hadron distribution amplitudes (DA), which contain all the non-perturbative physics.

The asymptotic distribution amplitude for the proton leads to $G_M^P(|Q^2|)=0$ for all $Q^2$, and many different variations of asymmetric DA's have been considered, with and without Sudakov corrections, and with and without transverse momenta.  QCD sum--rule predictions, and predictions based on GPD and meson--cloud pictures have also been made.  It is not surprising that with an appropriate choice of the parameters, the spacelike form factors of the proton can be fitted by nearly all model calculations.

\subsection{Timelike Form Factors of the Proton}

Prior to the Fermilab (E760/E835) measurements in 1993/2000 [8,9,10] of the timelike form factors of the proton by the reaction $p\bar{p}\to e^+e^-$, the data were sparse, had large errors, and were confined to $|Q^2|<5,7~\mathrm{GeV}^2$.  The Fermilab measurements \cite{e760,e835} obtained $G^p_M(|Q^2|)$ for four $|Q^2|$ between 8.9 and 13.11 $\mathrm{GeV}^2$.  As Fig.~1~(left) shows, while $Q^4G^p_M(|Q^2|)$ was found to vary as $\alpha^2(\mathrm{strong})$, the value of the timelike form factor was found to be twice as large as the spacelike form factor, i.e., 
$$R\equiv G^p_M(|Q^2|)(\mathrm{timelike})/G^p_M(|Q^2|)(\mathrm{spacelike})\approx2$$

Prior to the Fermilab measurements there were few theoretical predictions of the timelike form factor of the proton.  Following the Fermilab measurements, Hyer \cite{hyer} reported predictions for timelike form factors within the pQCD formalism including Sudakov suppression.  Hyer's predictions, showed large  sensitivity to the assumed distribution amplitude, but did not address the question of the experimental ratio $R\approx2$.  Iachello and Wan \cite{iachello} gave made predictions based on a picture of the bare meson surrounded by a vector meson cloud.  As shown in Fig.~1~(right), a typical Hyer prediction gives a $Q^4G_M^p(|Q^2|)$ which is nearly constant with $|Q^2|$, and the Iachello prediction gives $Q^4G_M^p(|Q^2|)$ which falls rapidly with $|Q^2|$.  Neither fits the data.

In order to explain the ratio $R\approx2$  Kroll and collaborators \cite{diquark}  proposed the diquark--quark model of the nucleon.  While this model has at least two extra parameters, it explains both spacelike and timelike $G^p_M(|Q^2|)$, and $R\approx2$ very nicely.  The three predictions mentioned above are shown in Fig.~1~(right), where the more recent results for $G_M^p(|Q^2|)$ from the $e^+e^-$ annihilation measurments by Cornell~\cite{cleo-ff} at $|Q^2|=13.5~\mathrm{GeV}^2$, by BES~\cite{bes}  at ten values of $|Q^2|=4-9.4~\mathrm{GeV}^2$, and by BaBar~\cite{babar} using ISR from $\Upsilon(4S)$ at $|Q^2|=3.6-20.3~\mathrm{GeV}^2$ are also shown.  All these measurements are consistent with each other, and confirm $R\approx2$.  BaBar has gone a step beyond, and has also attempted to derive $G_E/G_M$ from  their ISR data.

\begin{figure}[!tb]
\begin{center}
\includegraphics[width=2.2in]{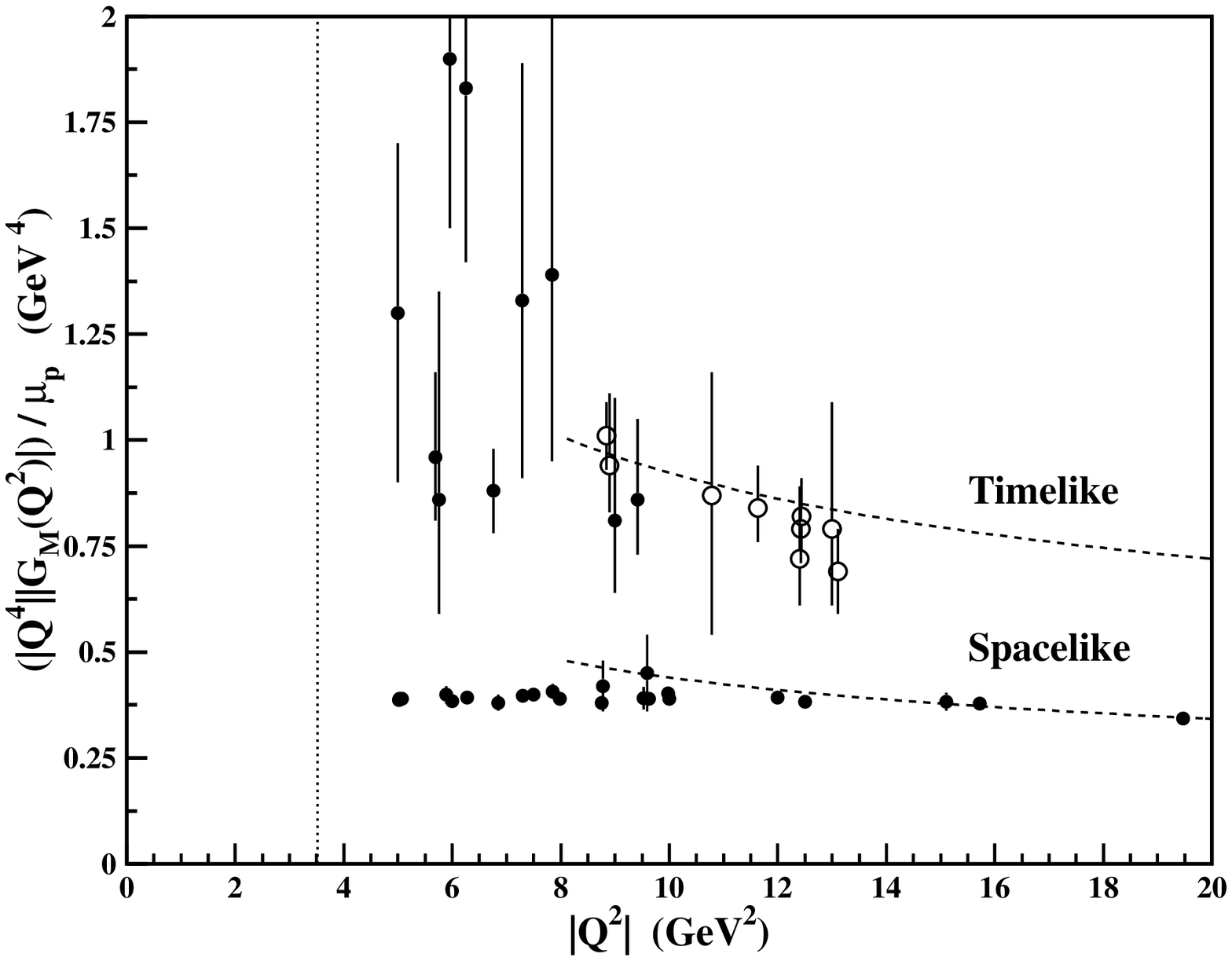}
\includegraphics[width=2.2in]{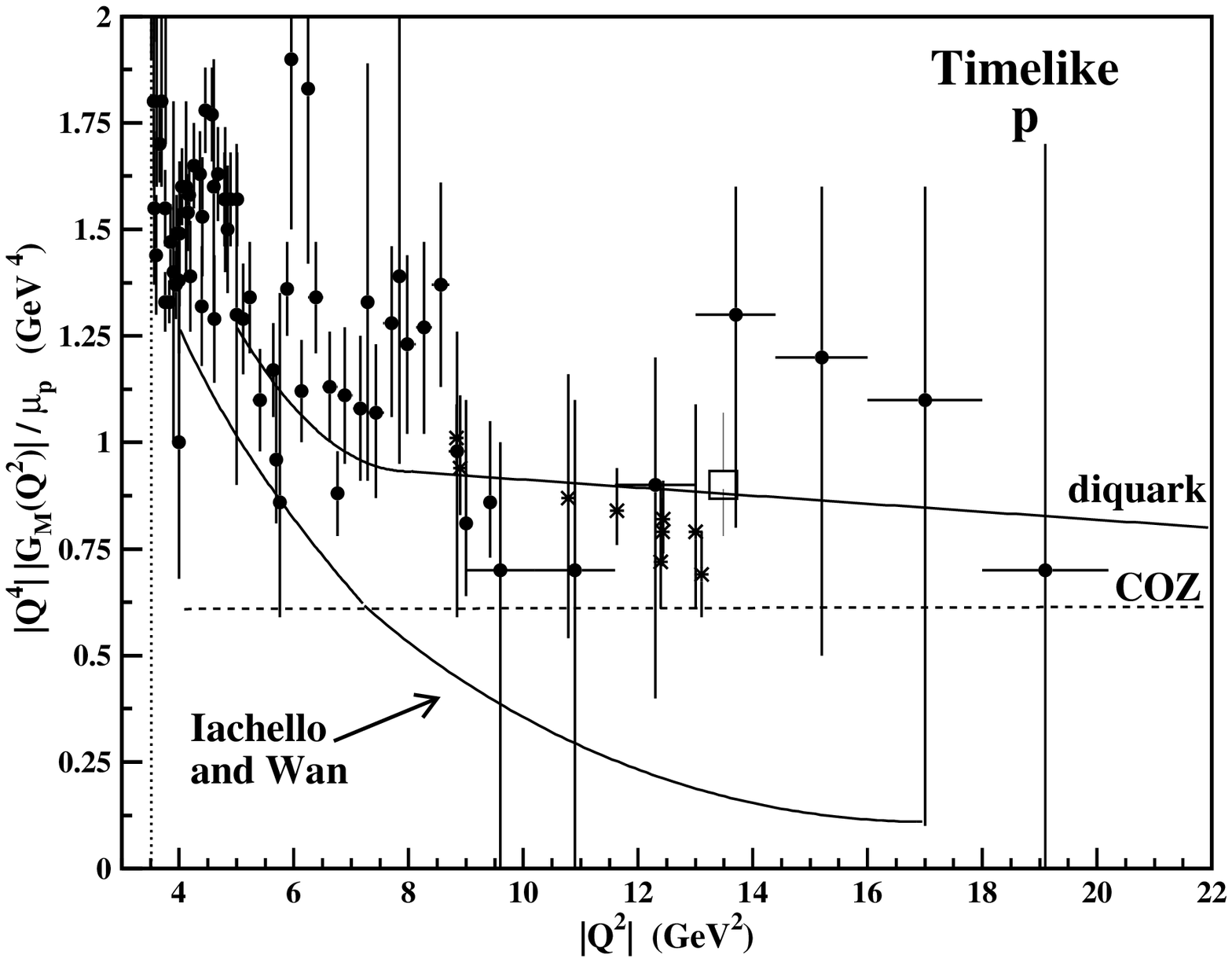}
\end{center}
\caption{$Q^4G^p_M(|Q^2|)/\mu_p$ as a function of $|Q^2|$ for timelike momentum transfers.  (Left): Illustrating the nearly factor 2 larger values of timelike $|Q^2|$ as measured in the Fermilab experiments. (Right) A sample of theoretical predictions together with all the present experimental data from Fermilab, CLEO, BES, and BaBar.}
\end{figure}

Before I leave the proton, and go on to the pion and kaon, let me point out some possible consequences of the recent JLab measurements \cite{jlab} of $R(|Q^2|)\equiv\mu_pG_E(|Q^2|)/G_M(|Q^2|)$ for spacelike $|Q^2|<6$~GeV$^2$.  As is well known, these polarization measurements show that $R$ decreases monotonically as $|Q^2|$ increases.  If this trend is extrapolated, one reaches $R\approx0$ at $|Q^2|\approx8$~GeV$^2$, and $R$ becomes negative for larger $|Q^2|$, e.g., $R\equiv-0.8$ at $|Q|^2=13.5$~GeV$^2$.  I do not know what zero and negative $G_E(|Q^2|)$ mean, but I am tempted to speculate about what these would imply for timeilike $R(|Q^2|)$.  


Strangely enough, if we assume that the ratio of the Pauli and Dirac form factors, $F_2(|Q^2|)/F_1(|Q^2|)$, is the same for timelike $|Q^2|$ as it is for spacelike, for $|Q^2|=13.5$~GeV$^2$ we obtain almost the same result, $[\mu_pG_E(13.5)/G_M(13.5)]_{\mathrm{timelike}}=3.9$~and~4.9, whether $R(13.5))_{\mathrm{spacelike}}=1$~or~$-0.8$.


\begin{figure}[!tb]
\begin{center}
\includegraphics[width=2.2in]{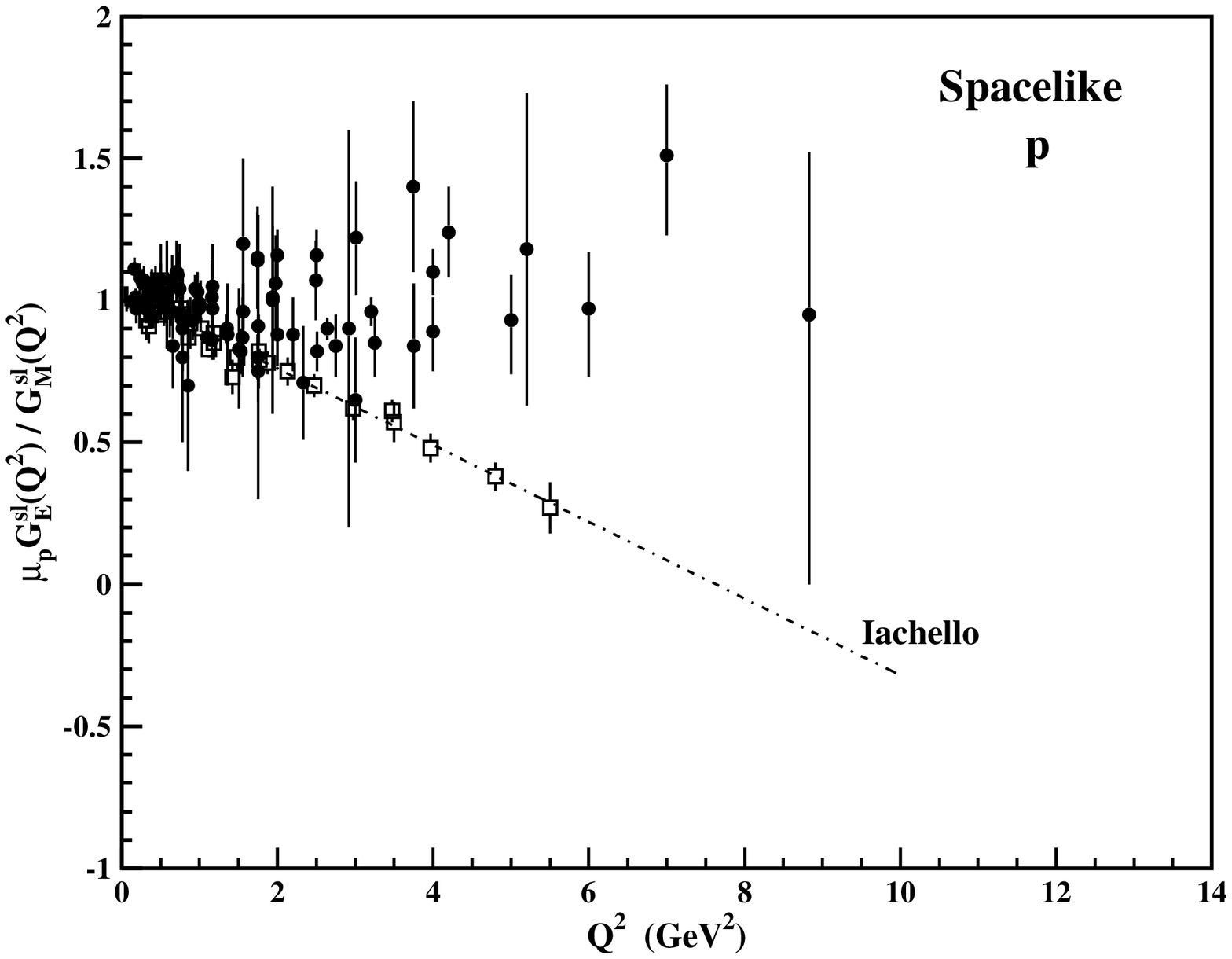}
\includegraphics[width=2.2in]{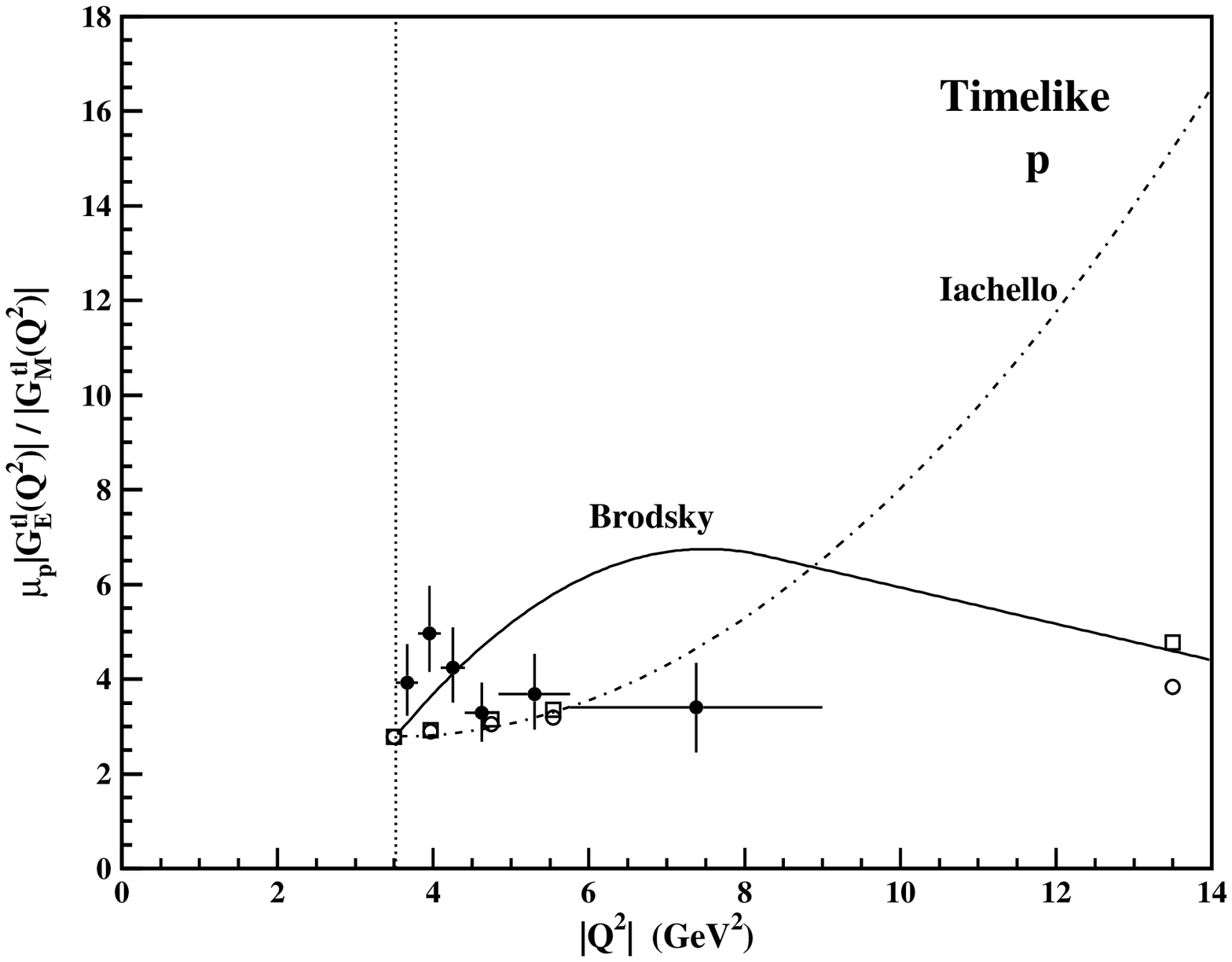}
\end{center}
\caption{(Left) $R\equiv\mu_pG^p_E(|Q^2|)/G^p_M(|Q^2|)$ for spacelike momentum transfers, \hspace*{1.cm} (Right) $R$ for timelike momentum transfers.  See text.}
\end{figure}


\section{Meson Form Factors}

Mesons represent much simpler systems than baryons; two quark systems are expected to be easier to understand than three quark systems.  It is because of this that the now-classic debate about when $|Q^2|$ is large enough for the validity of pQCD took place in the 1980s between Brodsky and collaborators on one side and Isgur and Llwellyn Smith on the other side.  Unfortunately, the then existing experimental data on pion form factors was extremely poor, especially in the large $|Q^2|$ region which was the subject of the entire debate.  All the pion and kaon form factor data available before the recent CLEO measurements are shown in Fig.~3.

\begin{figure}[!tb]
\begin{center}
\includegraphics[width=2.2in]{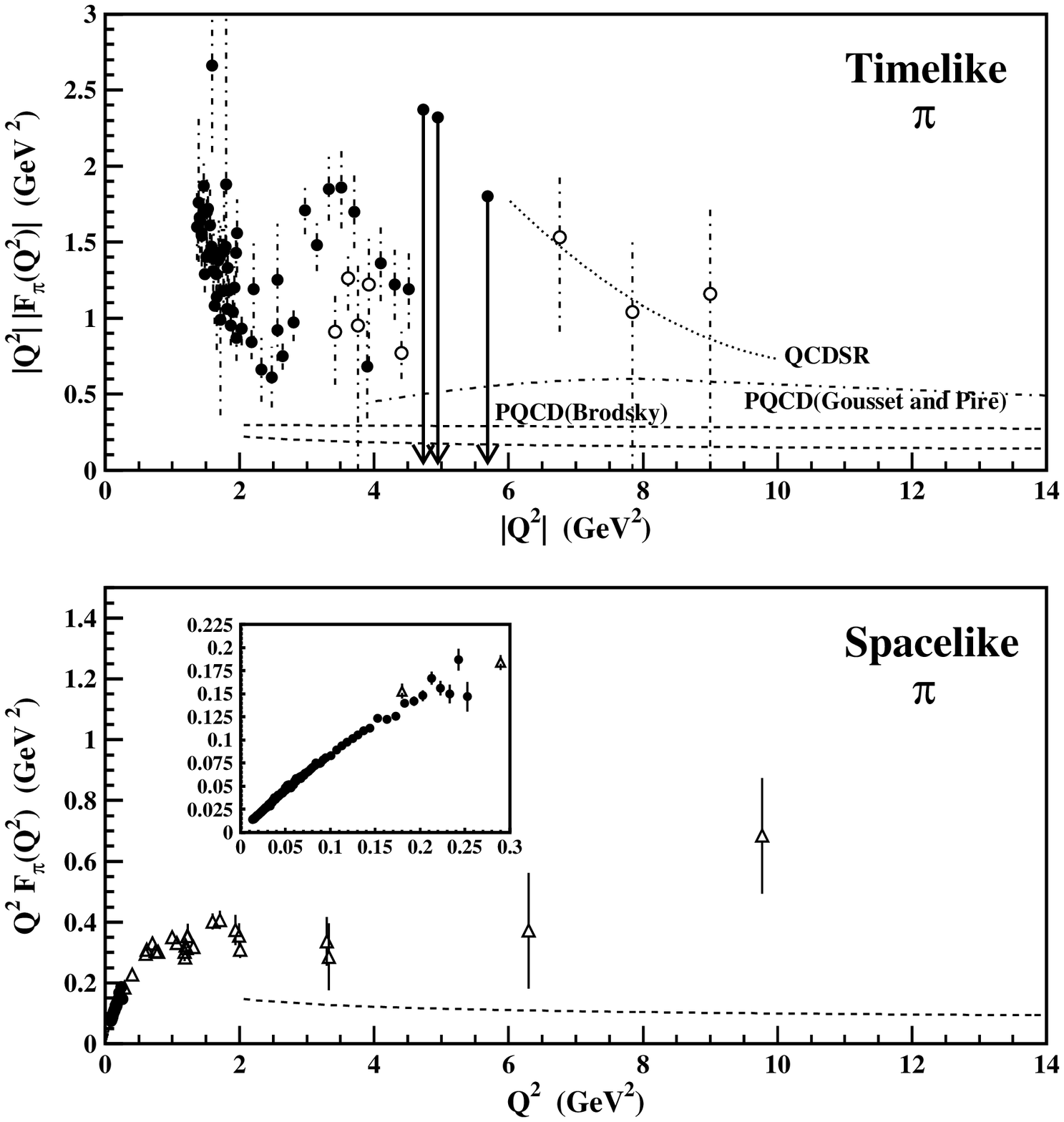}
\includegraphics[width=2.2in]{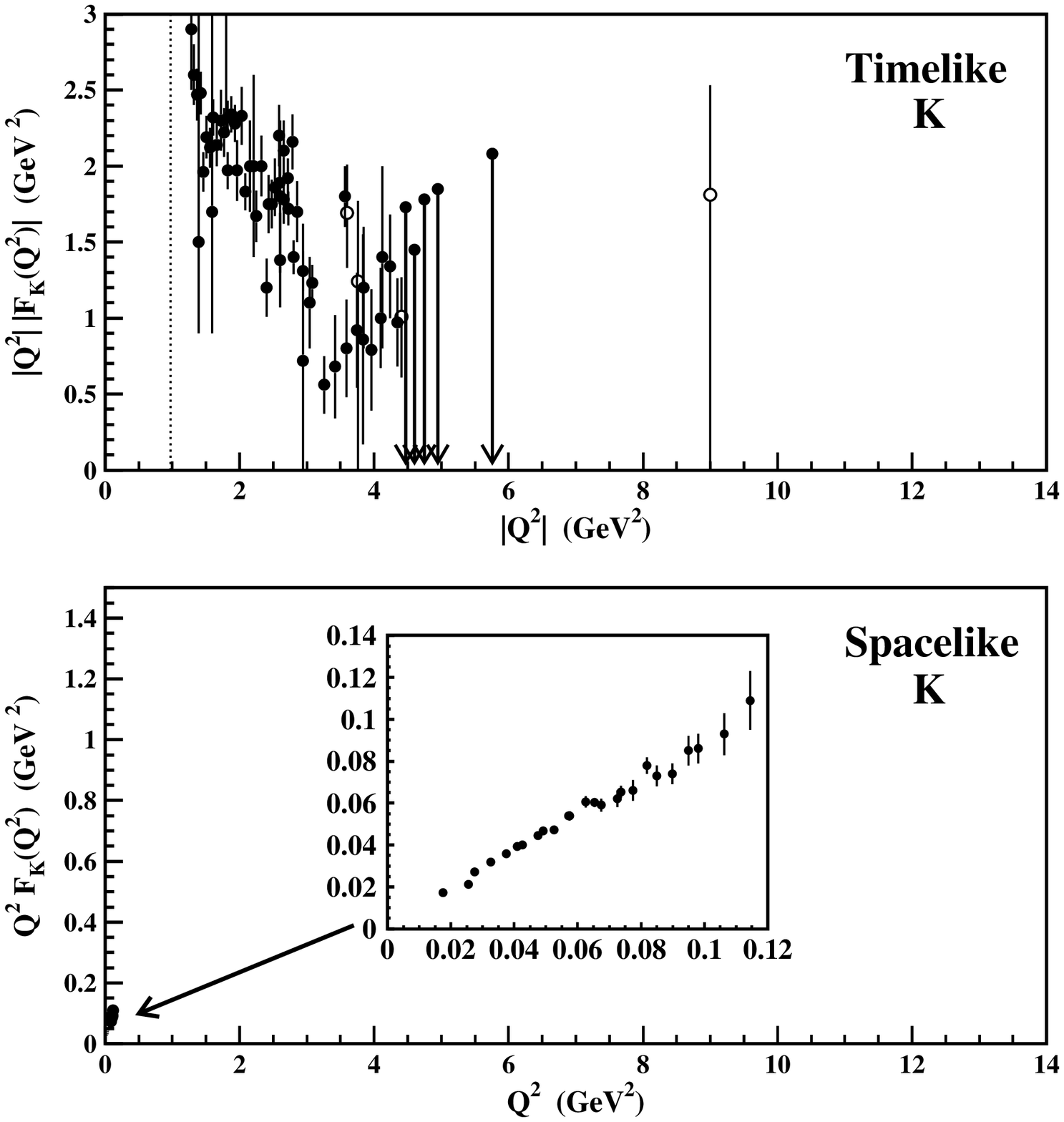}
\end{center}
\caption{World data for form factors for spacelike and timelike momentum transfers for pions (left) and kaons (right) before the CLEO measurements.}
\end{figure}

\subsection{Pion Form Factors}

Spacelike form factors of pions for $Q^2>0.3$~GeV$^2$ are exclusively determined by electroproduction measurements, $e^-p\to e^-\pi^-p,~e^-\pi^+n$.   The longitudinal part of the pion electroproduction cross section $\sigma_L(t)$ is related to $F^2_\pi(|Q^2|)_{\mathrm{spacelike}}$ via the pion--nucleon coupling constant $g_{\pi NN}(t)$.  The latest JLab electroproduction measurements \cite{jlab2} for $|Q^2|=0.6-2.45$~GeV$^2$ make the longitudinal/transverse separation for the first time, but still suffer from the uncertainties inherent in the $t$ dependence of $g_{\pi NN}(t)$ and the need to extrapolate the cross section to the physical pion pole at $t=m_\pi^2$.   The old larger $Q^2$ measurements from Cornell have the additional problem of very large ($\gtrsim\pm40\%$) errors.  Despite these limitations, the spacelike form factor data have been fitted by many model calculations with suitable choices of parameters.  For timelike form factors of the pion, the available data were sparse and generally of poor quality.  

\subsection{Kaon Form Factors}

For kaon spacelike form factors, there are no electroproduction measurements so far, and the available data are limited to $Q^2<0.12$~GeV$^2$.  The data for timelike form factors of kaons had the same limitations, both in the quality and the range of $Q^2$ as for pions.  There were no direct theoretical predictions, except that in the lowest order, it is expected that for all timelike $|Q^2|$, $F_\pi(Q^2)/F_K(Q^2)=f_\pi^2/f_K^2=0.67\pm0.01$.

\vspace*{-10pt}

\subsection{The CLEO Measurements of Pion and Kaon Form Factors}

CLEO \cite{cleo-ff} has recently reported measurements of the pion and kaon form factors for the timelike momentum transfer of $|Q^2|=13.48$~GeV$^2$.  Precision at the level of $\pm6\%$ for kaons, and $\pm13\%$ for pions has been achieved.  This unprecedented level of precision for a large $|Q^2|$ provides for the first time data which present a serious challenge to the theorists.

CLEO measurements of $\pi$ and $K$ form factors presented formidable background problems.  The form factor cross sections $\sigma(e^+e^-\to m\overline{m})\approx5-10$~pb, while $\sigma(e^+e^-\to e^+e^-,~\mu^+\mu^-)$ are $10^3$ to $10^5$ times larger.  Therefore, in addition to the standard track and shower quality requirements, very clever use of $dE/dx$, $E_{CC}$, and RICH information was done in order to obtain small but background free samples of $\pi^+\pi^-$ and $K^+K^-$ events.

The CLEO measurements were made  using 20.7 pb$^{-1}$ of $e^+e^-$ data taken at $\sqrt{s}=3.671~\mathrm{GeV}$, i.e., 15 MeV below the $\psi'$ resonance.  The data were originally taken for background studies for the $\psi'$ decays which were being studied.  It is ironic that these background studies have provided the world's best measurements of pion and kaon form factors.

\begin{figure}[!tb]
\begin{center}
\includegraphics[width=3.2in]{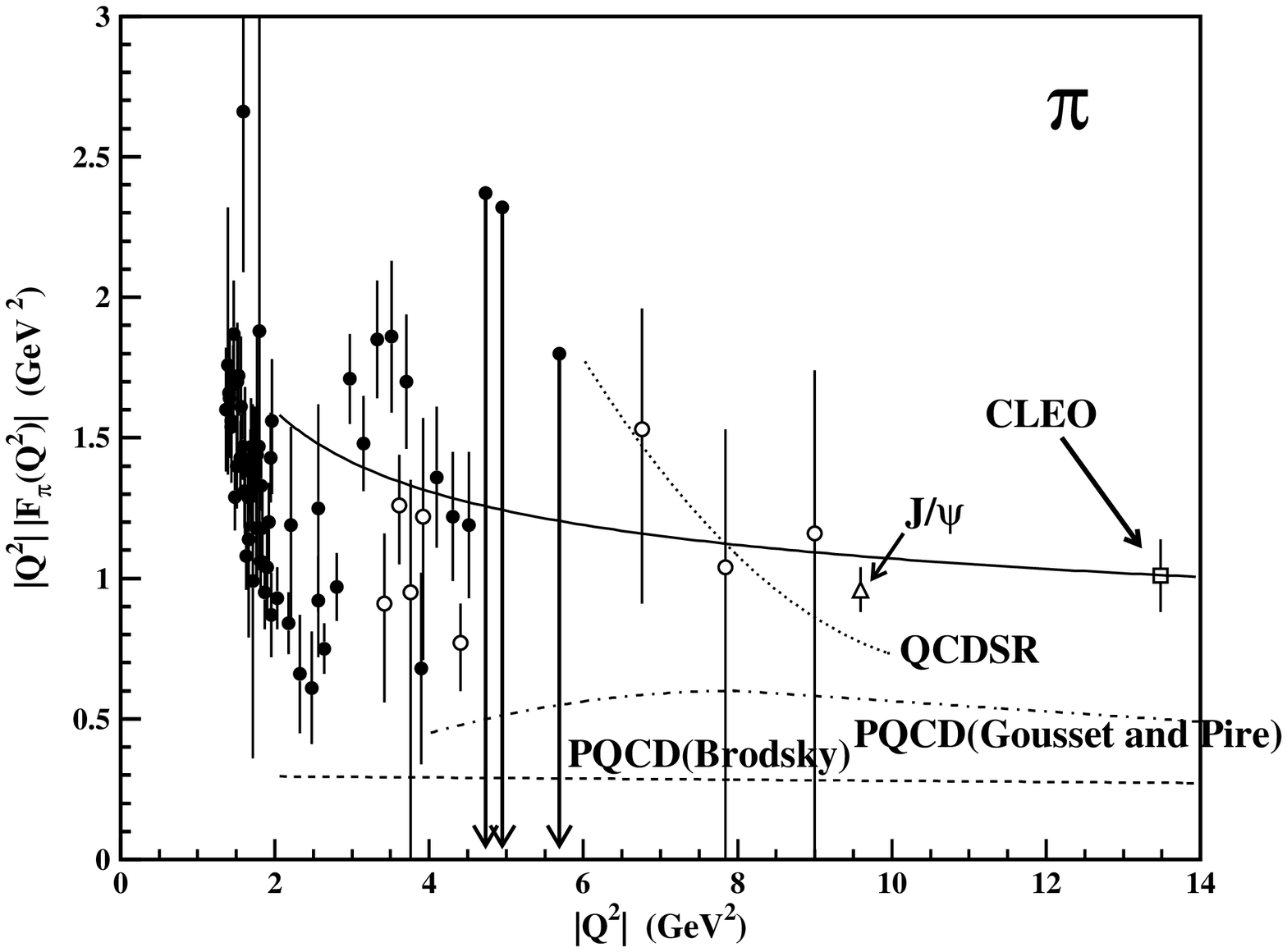}
\includegraphics[width=3.2in]{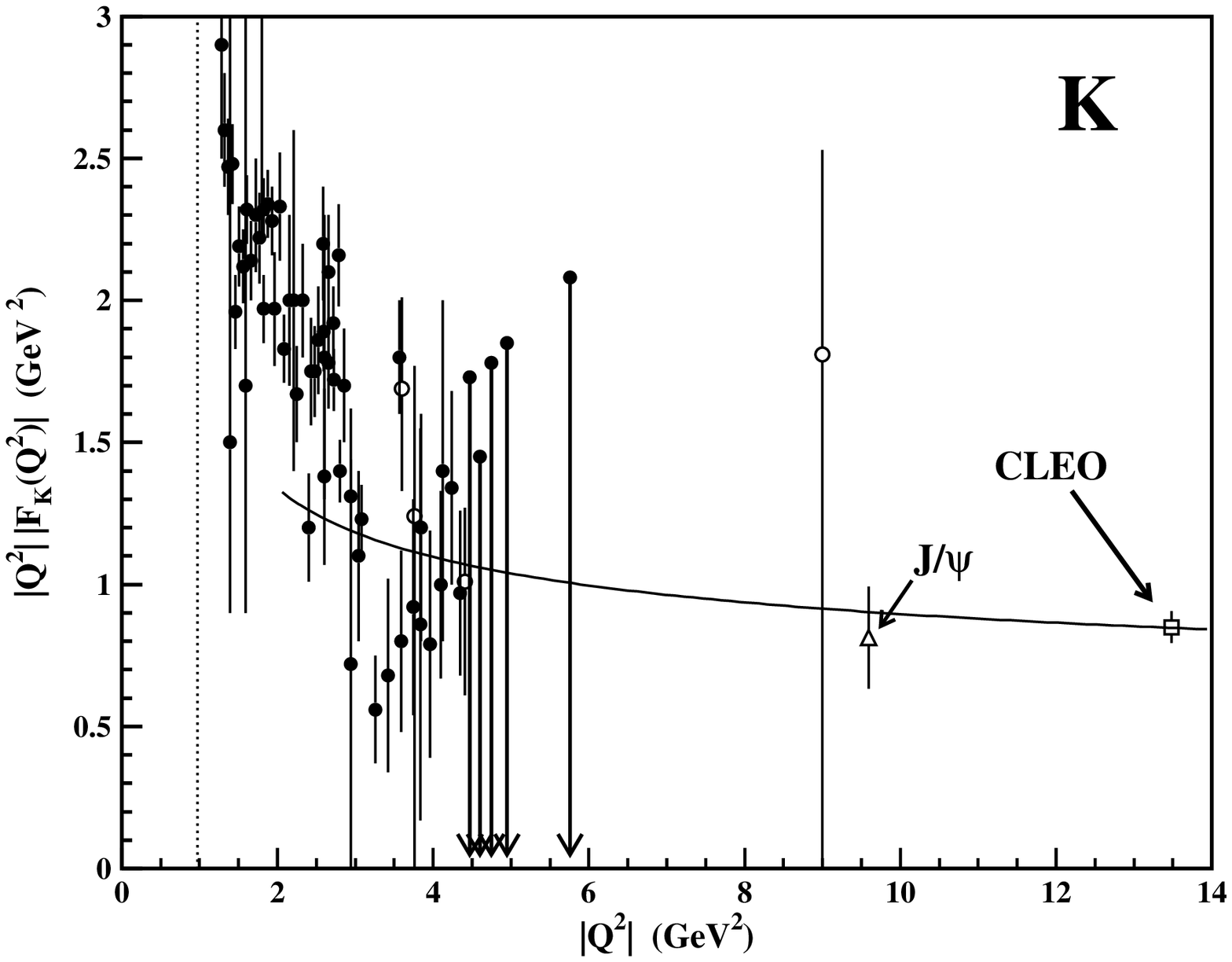}
\end{center}
\caption{World data for timelike form factors, including CLEO and $J/\psi$ results for pions (top) and kaons (bottom).  A sample of theoretical predictions available for pions is shown in addition to the arbitrarily normalized pQCD prediction of $\alpha_S$ variation (solid lines).}
\end{figure}

These data were analyzed to obtain for $|Q^2|=13.48~\mathrm{GeV}^2$: $|Q^2|F_\pi(|Q^2|)=1.01\pm0.11\pm0.07~\mathrm{GeV}^2$, $|Q^2|F_K(|Q^2|)=0.85\pm0.05\pm0.02~\mathrm{GeV}^2$, and $F_\pi(13.48~\mathrm{GeV}^2)/F_K(13.48~\mathrm{GeV}^2)=1.19\pm0.07$.

These are the world's first measurements of the form factors of any mesons at this large a momentum transfer, and with precision of this level.  They are shown in Fig.~5 along with the old world data, and arbitrarily normalized curves showing the pQCD predicted variation of $|Q^2|F_\pi$ and $|Q^2|F_K$ with $\alpha_S$.

In the figures for both pions and kaons, in addition to the results of the CLEO measurements  points marked $J/\psi$ are shown.  These have been obtained by using the relation
$$\frac{\mathcal{B}(J/\psi\to\gamma^*\to m\overline{m})}{J/\psi\to\gamma^*\to e^+e^-} = 2 F_m^2(M^2_{J/\psi})\times\left(\frac{p_m}{M_{J/\psi}}\right)^3$$
In general $\mathcal{B}(J/\psi\to m\overline{m})=K|A_\gamma + A_{ggg} + A_{\gamma gg}|^2$.  It was noted by Milana et al. \cite{milana} that for $\pi^+\pi^-$, $A_{ggg}$ and $A_{\gamma gg}$ are negligably small, so that $\mathcal{B}(J/\psi\to\gamma^*\to \pi^+\pi^-)=\mathcal{B}(J/\psi\to \pi^+\pi^-)$.  This was extended by us \cite{seth} to $K^+K^-$ by noting that
$$\mathcal{B}(J/\psi\to\gamma^*\to K^+K^-)\approx\mathcal{B}(J/\psi\to K^+K^-)-\mathcal{B}(J/\psi\to K_SK_L)$$
Using literature values for the branching fractions, the results are\\
$|Q^2|F_K(9.6~\mathrm{GeV}^2)=0.81\pm0.06$~GeV$^2$,\\
$|Q^2|F_\pi(9.6~\mathrm{GeV}^2)=1.01\pm0.13$~GeV$^2$,\\
both of which are in good agreement with the CLEO measurements for $|Q^2|=13.5$~GeV$^2$.  The ratio $F_\pi/F_K=1.19\pm0.07$ and $1.16\pm0,27$ for  $|Q^2|=13.5$~GeV$^2$ and  $|Q^2|=9.6$~GeV$^2$, respectively, which are both in disagreement with the pQCD expectation that $F_\pi/F_K=f_\pi^2/f_K^2=0.67\pm0.01$.

To summarize, for the first time we now have precision results for the form factors of charged pions and kaons for large timelike momentum transfers of 9.6~GeV$^2$ and 13.5~GeV$^2$.  None of the theoretical calculations, which exist only for pions, come even close to the experimental results.  In absence of precsion experimental results this could be tolerated.  Now there is no excuse.  The theorists must now go back to work on new QCD--based models for form factors which are among the most important measures of hadron structure.  It is interesting to note that no help is expected in this endeavour from Lattice practitioners who work in Euclidean time.  For the experimentalists, the challenge is to extend the precision measurements to as large momentum transfers as possible, and to other mesons and baryons.  BES III is ideally placed to meet this challenge.


\begin{thebibliography}{99}

\bibitem{slac1} A. F. Sill et al., Phys. Rev. \textbf{D48}, 29 (1993).

\bibitem{brodsky} G. P. Lepage and S. J. Brodsky, Phys. Rev. \textbf{D22}, 2157 (1980).

\bibitem{e760} E760 Collaboration, T. A. Armstrong et al., Phys. Rev. Lett. \textbf{70}, 121 (1993).

\bibitem{e835} E835 Collaboration, M. Ambrogiani et al., Phys. Rev. \textbf{D60}, 032002 (1999); E835 Collaboration, M. Andreotti et al., Phys. Lett. \textbf{B559}, 20 (2003).

\bibitem{hyer}  T. Hyer, Phys. Rev. \textbf{D47}, 3875 (1993).

\bibitem{iachello} F. Iachello and Q. Wan, Phys. Rev. \textbf{C 69}, 055204 (2004).

\bibitem{diquark} P. Kroll et al., Phys. Lett \textbf{B 316}, 546 (1993).

\bibitem{cleo-ff} CLEO Collaboration, T. Pedlar et al., Phys. Rev. Lett. \textbf{95}, 261803 (2005).

\bibitem{bes} BES Collaboration, M. Ablikim et al., Phys. Lett. \textbf{B630}, 14 (2003).

\bibitem{babar} BaBar Collaboration, B. Aubert et al., Phys. Rev. \textbf{D73}, 012005 (2006).

\bibitem{jlab} M. K. Jones et al., Phys. Rev. Lett. \textbf{84}, 1398 (2000); O.~Gayou~et~al., Phys.~Rev.~Lett.~\textbf{88}, 092301 (2002).

\bibitem{jlab2} J. Volmer et al., Phys. Rev. Lett. \textbf{86}, 1713 (2001);  V. Tadevosyan et al., nucl-ex/0607007; T. Horn et al., Phys. Rev. Lett. \textbf{97}, 192001 (2006).


\bibitem{milana} J. Milana, S. Nussinov, and M. G. Olsson, Phys.~Rev.~Lett.~\textbf{71}, 2533 (1988).

\bibitem{seth} K. K. Seth, Phys. Rev. \textbf{D 75}, 017301 (2007).

\end{thebibliography}
\end{document}